\documentclass[12pt]{article}
\usepackage[dvips]{graphicx}
\if@twoside \oddsidemargin 26pt
 \else \oddsidemargin 12pt\evensidemargin 12pt
\fi
\newcommand{\be}{\begin{eqnarray}}
\newcommand{\ee}{\end{eqnarray}}
\begin{document}
%%%%%%%%%%%%%%%%%%%%%%%%%%%%%%%%%%%%%%%%%%%%%%%%%%%%%%%%%%%%%%%%%%%%%
\title{On Nonexistence of Magnetic Charge in Pure Yang-Mills Theories}
\author{A. Kovner, M. Lavelle and D. McMullan\\
\\
  {\normalsize\it Department of Mathematics and Statistics,}\\
  {\normalsize\it University of Plymouth}\\
  {\normalsize\it   Plymouth, PL4 8AA, UK}}
\maketitle
%====================================================================
\begin{abstract}
\noindent We prove that magnetic charge does not exist as a
physical observable on the physical Hilbert space of the pure
$SU(2)$ gauge theory. The abelian magnetic monopoles seen in
lattice simulations are then interpreted as artifacts of gauge
fixing. The apparent physical scaling properties of the monopole
density in the continuum limit observed on the lattice are
attributed to the correct scaling properties of physical objects -
magnetic vortices, as first argued by Greensite et.\,al. We show
that a local gauge transformation of a certain type \lq\lq
creates" abelian monopole-antimonopole pairs along magnetic
vortices. This gauge transformation exists in pure $SU(N)$ gauge
theory at any $N$.
\end{abstract}
\bigskip
%====================================================================
\newpage
Understanding the mechanism of confinement in QCD has kept many peoples busy for 30 years. At present there are
two basic schools of thought on the subject, both motivated in large measure by early works of 't~Hooft. One
school maintains that the area law of the Wilson loop, and thus confinement, comes about due to condensation of
magnetic vortices in the QCD vacuum. This physical idea was proposed in the early days of QCD \cite{nielsenolesen}
and later formalised in \cite{thooft1}. This approach has been dormant for quite a while with the exception of a
couple of works \cite{tomboulis}, \cite{kovner92}. Interest in it has been rekindled in the recent years in the
lattice community following the work of Greensite and collaborators \cite{greensite1}, \cite{greensite}.

The other idea is known by the name of dual superconductivity and goes back to \cite{thooft2}. Here one assumes
that the QCD vacuum behaves in a way similar to a superconductor but with electric and magnetic quantities
interchanged. In particular magnetic monopoles are supposed to be condensed in the QCD vacuum leading to the dual
Meissner effect and therefore linear confinement of colour charges.

Although on the face of it the status of the two approaches is
similar, in fact there is a profound difference between them.
Magnetic vortices in gauge theories are physical objects in the
sense that it is possible to define a gauge invariant vortex
creation operator. The expectation value of this operator then can
be used as a gauge invariant order parameter for confinement
\cite{thooft1} and it indeed has been shown analytically
\cite{vertexept} and numerically \cite{vortexep} that its
behaviour changes above deconfinement transition. As discussed in
\cite{kovner00} in pure Yang-Mills theories the VEV of this
operator probes the mode of realisation of the global magnetic
$Z_2$ symmetry. This symmetry is a little peculiar since its group
element is not an exponential of a volume integral of some charge,
but rather an exponential of a surface integral. This group
element is nothing but the fundamental Wilson loop taken over the
contour at spatial infinity\cite{kovner00}
\begin{equation}
M={\rm Tr}{\cal P}\exp\{ig\oint_{R\rightarrow\infty}dl_iA^i\}
\label{one}
\end{equation}

  On the other hand no
gauge invariant order parameter is known for monopoles in
nonabelian theories. Moreover, although the monopoles are supposed
to carry a conserved magnetic charge, no gauge invariant magnetic
charge operator has ever been constructed in pure gauge theories.

Nevertheless, the monopole condensation mechanism has been studied
extensively on the lattice. To define monopoles on the lattice one
has to fix a particular abelian gauge. This is achieved by
choosing an adjoint field $\chi=\chi^a\tau^a$ and diagonalising it
by a gauge transformation. Once this diagonalisation is achieved,
the residual gauge symmetry is $U(1)^{N-1}$. The locations of
abelian monopoles are then identified by measuring the abelian
magnetic flux corresponding to the residual gauge group. The flux
through an elementary plaquette is defined modulo $2\pi$. If the
total flux that emanates through all the faces of a three
dimensional cube is greater than $2\pi$, one ascribes a magnetic
monopole to this particular cube. The choice of the adjoint field
$\chi$ specifies the \lq\lq abelian projection". It is usually
taken as a component of the gauge field strength, as a Polyakov
loop on a finite lattice \cite{digiacomo}, or as the lowest
eigenfunction of the lattice Laplacian operator \cite{laplaciang}.
Any other choice is in principle permissible. The most popular
projection is the so called\lq\lq maximal abelian projection"
which is defined implicitly through minimisation of a particular
functional of lattice gauge fields \cite{suzuki}. The detailed
properties of the monopoles, such as their locations and density
clearly depend on the choice of the abelian projection
\cite{digiacomo}. Nevertheless it has been found that some basic
properties are common to monopoles defined in different abelian
gauges, at least within certain limits \cite{digiacomo}. For
example all monopoles are condensed in the confining phase and are
not condensed in the high temperature deconfined phase. Likewise
the monopole density in various gauges seems to scale similarly in
the continuum limit. This universality is at the root of the
frequently expressed belief that abelian monopoles are {\it bona
fide} physical objects \cite{zakharov} that carry a gauge
invariant conserved monopole charge \cite{zakharov1}.

This interpretation is frequently taken even further. One
postulates that the low energy dynamics of QCD is the same as that
of a dual abelian Higgs model, where the role of the Higgs field
is played by the monopole field and the role of the electric
charge by the magnetic charge. The lattice data is sometimes
interpreted in terms of this model \cite{polikarpov}. The dual
superconductor philosophy does not differentiate between the
$SU(N)$ gauge theories at $N=2$ and $N>2$. In practice this
approach is most frequently  applied at $N=2$, since this theory
is technically the simplest.

The purpose of this note is to show that an interpretation of this type is seriously flawed. In particular we will
show that the physical Hilbert space of pure $SU(2)$ gauge theory does not contain a physical observable which
could correspond to a magnetic charge. Therefore, at least in this case, the dual Higgs model cannot be the
correct low energy theory.

 The argument is very simple and straightforward. Consider pure $SU(2)$ gauge theory.
 Let us assume that within this theory there exists a local, gauge
 invariant magnetic current $J^M_\mu$. Such a current obviously
 must be $C$-parity odd, $C^{-1}J_\mu^MC=-J_\mu^M$. This is the case in all the
 implementations of the dual Higgs model known to us, and quite
 generally must be the case as monopoles have to be charge
 conjugates of anti-monopoles.
 However, it is a basic fact that  pure $SU(2)$ gauge
 theory {\it does not possess a physical charge conjugation
 transformation}. This is easiest to see in the lattice formulation.
 The basis of physical operators of a pure $SU(2)$ gauge theory is spanned
 by traces of fundamental Wilson loops of various sizes. The action of charge
 conjugation on any Wilson loop operator is
 \begin{equation}
 C^{-1}WC=W^*
 \label{C}
 \end{equation}
 However for  $SU(2)$ the group trace of any Wilson loop is real.
 Therefore any physical gauge invariant observable $O$ satisfies
\begin{equation}
 C^{-1}OC=O
 \end{equation}
Another way of seeing this, is to realise that since the two eigenvalues of $W$ are complex conjugates of each
other, the transformation  (\ref{C}) permutes the two eigenvalues. Thus the \lq\lq would be" charge conjugation is
nothing but the global Weyl subgroup of the $SU(2)$ gauge group.

Thus there are no charge conjugation odd observables in the pure
$SU(2)$ gauge theory, and therefore magnetic current and magnetic
charge do not exist. The $U(1)$ dual abelian Higgs model therefore
can not be a relevant low energy description of physical degrees
of freedom.

The preceding argument leaves the door open to the existence of a $Z_2$ subgroup of the monopole charge group. The
point is that even though a $U(1)$ charge is naturally $C$-odd, there is one element of the group generated by it
which is $C$-even. This group element $T=\exp\{\pi\int d^3x J^M_0\}$ could in principle exist\footnote{Note that
such an operator is \emph{not} the same as the $Z_2$ magnetic symmetry operator $M$ discussed in \cite{kovner00}.}
even if the charge itself does not. If such a $Z_2$ symmetry existed one could hope that the effective low energy
theory could be a $Z_2$ gauge Higgs model. However, this too can not be the case. A field that is odd under $T$
would be a source of  magnetic flux equal to half that of the minimal Dirac string. Thus a $Z_2$ magnetic vortex
could end on a particle created by such a field. However it is well known that only closed $Z_2$ vortices exist in
the $SU(2)$ gauge theory \cite{kovner00}. If open vortices existed, the magnetic symmetry group element $M$ of
 (\ref{one}) would not be conserved. We thus conclude that even if the low energy effective $Z_2$ gauge theory
existed, it could not contain matter fields \footnote{We note that such an effective theory has in fact been
suggested in \cite{tomboulis1}, but it has nothing whatsoever to do with the dual abelian Higgs model.}.

We note that our argument applies only to pure gauge theory and does not contradict the well established fact that
in the Georgi Glashow model ($SU(2)$ gauge theory with an adjoint Higgs field) one can perfectly well define a
gauge invariant magnetic charge. In this case the definition, due to 't Hooft, is
\begin{equation}
J_\mu^M=\partial_\nu \tilde F^{\mu\nu}
\label{magnc}
\end{equation}
with
\begin{equation}
F^{\mu\nu}=\hat\Phi^aF^a_{\mu\nu}-{1\over g}
f^{abc}\hat\Phi^a(D_\mu \hat\Phi)^b(D_\nu\hat\Phi)^c,\ \ \ \ \
\hat\Phi^a={\Phi^a\over |\Phi|}\,. \label{magnf}\end{equation}
where $\Phi^a$ is the adjoint Higgs field appearing in the
Lagrangian of the Georgi Glashow model. The charge conjugation
transformation in the Georgi Glashow model is defined as
\begin{eqnarray}
&&A^{1,3}_\mu\rightarrow -A^{1,3}_\mu\,, \ \ \ A^2_\mu\rightarrow
A^2_\mu\,,\nonumber\\
&&\Phi^{1,3}\rightarrow \Phi^{1,3}\,, \ \ \ \Phi^2\rightarrow
-\Phi^2\,.
\end{eqnarray}
This is clearly distinct from a global gauge transformation, since
the vector potential and the Higgs field transform differently.
However the transformation on the gauge potential alone is indeed
equivalent to a global gauge transformation. Thus when an
independent Higgs field is not present, the charge conjugation
also disappears. Physically this means that as the Higgs field is
made heavier, all the states in the Hilbert space which are odd
under the charge conjugation also become heavy. In the limit of
pure gluodynamics, or infinite Higgs mass, they become infinitely
heavy and therefore unphysical.

It is instructive to see how our argument coexists with the construction of a formally gauge invariant magnetic
current given in \cite{digiacomo}. The expression suggested in \cite{digiacomo} is the same as
eqs.(\ref{magnc},\ref{magnf}) with a composite field $\chi^a$ substituted for the Higgs field $\Phi^a$. The field
$\chi$ is chosen as some operator in the adjoint representation of the gauge group in gluodynamics and is the same
field which defines the abelian projection. The main problem with this expression, from our point of view, is that
it is not a Lorentz vector current, since the \lq\lq Higgs" which it is constructed from is not a scalar field (we
remind the reader that $\chi$ is usually chosen either as some fixed component of the gauge field strength, or as
a Polyakov loop on the finite lattice, or in some rather implicit way). Thus the spatial integral of the zeroth
component of this \lq\lq current" is not a scalar charge. It is indeed clear that this expression is charge
conjugation even, since the transformation properties of any adjoint $\chi$ under the \lq\lq charge conjugation"
(\ref{C}) are the same as those of $F^a_{\mu\nu}$. Choosing $\chi$ as a Lorentz scalar is not possible in $SU(2)$.
For example, the obvious choice $\chi^a=d^{abc}F^b_{\mu\nu}F^c_{\mu\nu}$ does not exist due to the vanishing of
the $d$-tensor in $SU(2)$. Thus the expression constructed in \cite{digiacomo} although gauge invariant, is not a
conserved scalar charge.

Another expression was suggested in \cite{zakharov1}. It is similar to the one in \cite{digiacomo}, except the
choice of $\chi$ is somewhat more intricate. This suggestion has an additional problem, namely that the abelian
gauge in \cite{zakharov1} is only fixed up to a discrete $Z_2$ gauge transformation - the Weyl subgroup of the
gauge group. The sign ambiguity in the definition of the magnetic charge noted in \cite{zakharov1} is precisely
the local version of the \lq\lq charge conjugation" gauge transformation which is central to  our argument. The
incomplete gauge fixing should, strictly speaking, force the charge to vanish in any finite region of space when
averaged over a gauge invariant ensemble \footnote{An explicit attempt at fixing the Weyl subgroup of the gauge
group was made in the lattice implementation in \cite{gubarev}. It leads to a nontrivial problem of finding a
global minimum of a certain Ising-type model.}.

Another definition of magnetic charge is used in the studies in
the maximal Abelian gauge. This gauge is defined by the
requirement that the following expression (we use here continuum
notations for simplicity) be globally minimized:
\begin{equation}
\int d^4x [(A_\mu^1)^2+(A_\mu^2)^2] \label{mag1}\end{equation}
The
magnetic monopole current is then defined as the divergence of the
third component of the dual field strength,
\begin{equation}
j_\mu^{MAG}=\partial_\nu\tilde F_{\mu\nu}^3
\label{mag}\end{equation} There is however the following obstacle
to this procedure. The MAG condition is itself invariant under the
discrete subgroup of the gauge group that is the central point of
our discussion. It is clearly invariant under the change of sign
of $A^1$ and $A^3$, which is generated by the transformation
eq.(\ref{C}). Since this part of the gauge group is not fixed, in
any properly generated ensemble of field configurations, for any
given configuration one must find a gauge copy which differs only
by this transformation. The definition of magnetic current in
eq.(\ref{mag}) is odd under the transformation in question. Thus
when averaged over the gauge copies, the current thus defined
vanishes. This is another way of saying that this current is not
gauge invariant.

 Our proof of nonexistence of magnetic charge seems to
lead to an apparent paradox. As we have mentioned above, lattice simulations of the pure $SU(2)$ gluodynamics do
find abelian monopoles. The density of these monopoles is gauge dependent, but it nevertheless seems to scale
correctly in the continuum limit. How can this be the case if the monopoles themselves are not physical objects?
We believe that the correct answer to this question is the one put forward in \cite{greensite}. According to
\cite{greensite}, the abelian magnetic monopoles that are found in lattice simulations overwhelmingly reside on
magnetic vortices. The magnetic vortices are in fact physical objects and, being condensed in the vacuum, their
density must scale in the continuum limit\footnote{One has to be cautious with the interpretation of this
statement in the following sense. While the density of physical, gauge invariant magnetic $Z_2$ vortices must
scale correctly in the continuum limit, the existing lattice algorithms for identifying magnetic vortices are
themselves not gauge invariant. Accordingly in different gauges their density has different scaling properties.
For example, in the maximum center gauge the density scales \cite{vortexscaling} while in the Laplacian center
gauge it does not \cite{laplacianvortex}. The relevant question is how well a given \lq\lq vortex finding
algorithm" identifies physical vortices. Indirect tests performed in \cite{greensite1} suggest that the maximal
center gauge does the job rather well. At any rate, since the explicit expression for the conserved $Z_2$ charge
carried by the vortices exists\cite{kovner00}, one should be able to set up an explicitly gauge invariant
procedure for identifying physical vortices on the lattice. This would involve measuring the sign of Wilson loops
of various sizes, and identifying  the underlying vortex structure on the basis of this data. Such a procedure
however may turn out to be tricky since it may be affected by  ultraviolet lattice artifacts. We thank Philippe de
Forcrand for raising this issue with us.}. The positions of monopoles along the vortex are determined by the
specific abelian projection and are more or less random. Thus although the monopoles are gauge artifacts
themselves, their locations trace the locations of physical objects --- magnetic vortices --- and therefore scale
in the continuum limit. The numerical evidence presented in \cite{greensite} is very supportive of this point of
view. Not only were monopoles found to reside on vortices, but also  the distribution of the energy density around
a monopole was found to be very similar to that around any other point on the vortex. In fact in a particular
(center Laplacian) gauge it was proven in \cite{deforcrand} that monopoles are {\it always} located on vortices.

In the rest of this note we want to demonstrate that there is a clear connection between the local version of the
gauge transformation (\ref{C}), on which our argument hinges, and the fact that monopoles are located along
vortices. Consider a lattice $SU(2)$ pure gauge theory in an abelian gauge specified by diagonalisation of some
adjoint operator $\chi(x)$ defined on lattice sites. In this abelian projection one defines vortices and monopoles
in terms of a magnetic flux in the third direction in colour space. The abelian part $P_A$ of every $SU(2)$
plaquette matrix $P$ is then defined through the Euler angle decomposition~\cite{digiacomo}
\begin{eqnarray}
P(x)&=&\exp\{i\alpha(x)\sigma_3\}\exp\{i\beta(x)\sigma_2\}\exp\{i\gamma(x)\sigma_3\}
\\
&=& \exp\{i\alpha(x)\sigma_3\}\exp\{i\beta(x)\sigma_2\}\exp\{-i\alpha(x)\sigma_3\} P_A(x)\,,\nonumber
\end{eqnarray}
with
\begin{equation}
P_A(x)=\exp\{i(\gamma(x)-\alpha(x))\sigma_3\}\equiv\exp\{iF(x)\sigma_3\}
\end{equation}
where $\sigma_i$ are the  Pauli matrices. Here the plaquette variable is defined in a standard way as a product of
the $SU(2)$ link matrices along the plaquette:
\begin{equation}
P^{ab}_{\mu\nu}(x)=[U(x,x+\mu)U(x+\mu,x+\mu+\nu) U(x+\mu+\nu,x+\nu)U(x+\nu,x)]^{ab}\,.
\end{equation}
Let us now consider a local version of the gauge transformation (\ref{C}), which permutes the eigenvalues of the
matrix $\chi(x)$ at one particular site $X$. This transformation is affected by the matrix $c(x)$ which is unity
on all sites except for $X$, and on this site is given by $\sigma_2$. Under this transformation the plaquette
matrix transforms as
\begin{equation}
P_{\mu\nu}(x)=c^{-1}(x)P_{\mu\nu}(x)c(x)\,.
\end{equation}
Thus the only plaquettes that are affected by this transformation are the ones that are associated with the site
$X$. For all these plaquettes (i.e., all orientations $\mu\nu$) the appropriate abelian parts are conjugated by
this transformation
\begin{eqnarray}
c^{-1}P_{A,\mu\nu}(X)c&=&P^*_{A,\mu\nu}(X)\,,\nonumber\\
c^{-1}F_{\mu\nu}(X)c&=&-F_{\mu\nu}(X)\,.
\end{eqnarray}
Now, if the abelian field $F_{\mu\nu}(X)$ is small this transformation has no particular significance. However if
the gauge field configuration is such that a $Z_2$ magnetic vortex is piercing one of the plaquettes associated
with the point $X$, the abelian magnetic field on this plaquette is close to $\pi$. The gauge transformation then
transforms the flux on this particular plaquette from $\pi$ to $-\pi$ thus reversing the direction of the magnetic
flux of the vortex on this particular plaquette. This of course appears  equivalent to placing a
monopole-antimonopole pair in the three dimensional cubes separated by this plaquette. Clearly one can perform a
gauge transformation not on one site but on several sites along the magnetic vortex. A transformation of this type
will place on the vortex   monopole-antimonopole pairs separated by arbitrary distances. We stress again that
plaquettes not pierced by a magnetic vortex do not undergo any major change as a result of this transformation.

We have thus established that the local version of the gauge
transformation (\ref{C}) places monopole-antimonopole pairs along
$Z_2$ magnetic vortices. In a sense, it transforms a segment of a
vortex into an antivortex. Since charge conjugation is not a
physical symmetry in this system, a vortex and an antivortex are
physically equivalent and thus any monopole-antimonopole pair that
distinguishes between them is a pure gauge artifact. This goes
nicely together with the finding of \cite{greensite} that in terms
of energy density monopoles are hardly at all distinguishable from
other points along vortices.

Our proof of nonexistence of the magnetic charge applies strictly
speaking only to the $SU(2)$ theory. Gauge theories with higher
$SU(N)$ gauge groups do possess a physical charge conjugation
symmetry and thus in those theories a conserved magnetic current
can be constructed. In particular the simplest candidate for such
a current would be eqs.(\ref{magnc},\ref{magnf}) with the Higgs
field chosen as a bona fide scalar
$\chi^a=d^{abc}F^b_{\mu\nu}F^c_{\mu\nu}$. However, even in this
case, the naive dual abelian Higgs model is not a viable low
energy theory, since it is supposed to involve $N-1$ magnetic
charges. Some of these currents are clearly unphysical. In
particular the analogue  of the local gauge transformation that
artificially creates monopole-antimonopole pairs clearly exists in
any pure $SU(N)$ gauge theory. In this case such a segment of a
$Z_N$ magnetic vortex can be transformed into $N-1$ antivortices,
and the two are physically equivalent for any $N$. Thus at least
some monopoles in the lattice simulations of $SU(3)$ gauge
theories are also gauge artifacts. There are also other
configurations in the $SU(3)$ gauge theory which are identified as
monopoles by the existing \lq\lq monopole finding" algorithms. In
particular one can imagine three $Z_3$ magnetic vortices coming
together at the same point $X$. Such a configuration clearly is
not equivalent to a single vortex, but rather to two vortices
which run together along the same line and, at the point $X$,
split into two. These configurations must exist in lattice
simulations. The question is quantitative: whether their density
can be accounted for simply by the probability of two independent
vortices running along the same line, or is this density
significantly enhanced? Of course even these configurations do not
look like monopoles with a Coulomb-like magnetic field. Whether
there are genuine physical Coulomb-like monopoles in this case is
an interesting question. It seems unlikely to us that this is the
case, but the question is certainly well worth studying
numerically. In particular, it would be interesting to extend the
numerical analysis of \cite{greensite} to the $SU(3)$ theory and
see if a statistically significant proportion of monopoles exists
for which the fluxes through each phase of the cube are different
from an integer multiple of $2\pi/3$ modulo small fluctuations. It
would also be interesting to choose the abelian projection with
respect to the local scalar field
$\chi^a=d^{abc}F^b_{\mu\nu}F^c_{\mu\nu}$ and investigate
numerically whether properties of monopoles in this particular
gauge are any different from other gauges.

{\bf Note added.} The first version of this note was followed by the appearance of \cite{zakharov2}. This paper
summarises the view of its authors on the nature of abelian monopoles. It disagrees with our views on two main
points. First it suggests that the absence of $C$-parity is not an obstacle for definition of the monopole current
in maximal Abelian gauge. We have added a paragraph in the present version (around eqs.(\ref{mag1},\ref{mag}))
which shows that this is not the case, and that the current defined as in \cite{zakharov2} vanishes when averaged
over gauge copies.

The second important statement made in \cite{zakharov2} is that monopoles can not be gauge artifacts since the
action measured on the plaquettes bounding the monopole is greater (in lattice units) than the average plaquette
action on the lattice. In our view however this fact alone does not by itself preclude the monopoles from being
gauge fixing artifacts. As we have explained, we view monopoles as points on a magnetic vortex. The identification
of points on a vortex as monopoles depends on the gauge fixing and is thus a gauge fixing artifact. On the other
hand, a magnetic vortex being a physical object, must carry an excess of energy (or Euclidean action) at each of
its points relative to the average point on a lattice. Thus it is only natural that the points on the vortex that
are identified as monopoles by a particular lattice \lq\lq monopole finding algorithm" also carry an excess of
action. We do not claim that a monopole can be created by a gauge transformation at a point on the lattice which
has a small value of the field strength and thus our argument does not imply that they should be indistinguishable
from an \lq\lq average" point on the lattice.

\leftline{\bf Acknowledgements} We thank Jeff Greensite, Fedor Gubarev and Valya Zakharov for useful
correspondence and Philippe de Forcrand and Oliver Jahn for illuminating discussions and for pointing out
ref.\cite{vortexep}, \cite{laplaciang} and \cite{laplacianvortex} to us. We also thank Stephan Olejnik for
bringing ref.\cite{vortexscaling} to our attention. A.K. is supported by a PPARC advanced fellowship.


\begin{thebibliography}{99}
\vspace{-.5cm}
%\cite{Kovner:2001bh}


\bibitem{nielsenolesen} H.B.\ Nielsen and P.\ Olesen;
Nucl.Phys.B61:45-61,1973;\\ Nucl.Phys.B160:380,1979


\bibitem{thooft1} G.\ 't Hooft; Nucl.Phys.B138:1,1978



\bibitem{tomboulis} T.G.\ Kovacs and E.T.\ Tomboulis;
Phys.Lett.B321:75-79,1994 e-Print Archive: hep-lat/9311005

\bibitem{kovner92}A.\ Kovner and  B.\ Rosenstein;
Int.J.Mod.Phys.A7:7419-7514,1992; Int.J.Mod.Phys.A8:5575-5604,1993 e-Print Archive: hep-th/9212024


\bibitem{greensite1}  L.\ Del Debbio, M.\ Faber, J.\ Greensite  and
S.\ Olejnik; \\Phys.Rev.D55:2298-2306,1997 e-Print Archive: hep-lat/9610005

\bibitem{greensite}L.\ Del Debbio, M.\ Faber, J.\ Greensite  and S.\ Olejnik,
 In \lq\lq Zakopane 1997, New Developments in
Quantum Field Theory\rq\rq\ 47-64. e-Print Archive: hep-lat/9708023; J.\ Ambjorn, J.\ Giedt and J.\ Greensite;
JHEP 0002:033,2000 e-Print Archive: hep-lat/9907021

\bibitem{thooft2}G.\ 't Hooft; Nucl.Phys.B190:455,1981

\bibitem{vertexept}C.\ Korthals-Altes, A.\ Kovner and M.A.\ Stephanov; Phys.Lett.B469:205-212,1999;  e-Print Archive: hep-ph/9909516

\bibitem{vortexep}P.\ de Forcrand, M.\ D'Elia and M.\ Pepe; Phys.Rev.Lett.86:1438,2001 e-Print Archive:
hep-lat/0007034; P.\ de Forcrand and L.\ von Smekal;
Phys.Rev.D66:011504,2002 e-Print Archive: hep-lat/0107018

\bibitem{kovner00} C.\ Korthals-Altes and A.\ Kovner;
Phys.Rev.D62:096008,2000 e-Print Archive: hep-ph/0004052


\bibitem{digiacomo}  A.\ Di Giacomo, B.\ Lucini, L.\ Montesi and
G.\ Paffuti; Phys.Rev.D61:034503,2000 e-Print Archive: hep-lat/9906024; Phys.Rev.D61:034504,2000 e-Print Archive:
hep-lat/9906025;
 J.M.\ Carmona, M.\ D'Elia, A.\ Di Giacomo, B.\ Lucini and G.\ Paffuti;
Phys.Rev.D64:114507,2001 e-Print Archive: hep-lat/0103005

\bibitem{laplaciang} A.J.\ van der Sijs; Nucl.Phys.Proc.Suppl.53:535-537,1997;
e-Print Archive: hep-lat/9608041;
Nucl.Phys.Proc.Suppl.73:548-550,1999 e-Print Archive:
hep-lat/9809126



\bibitem{suzuki} T.\ Suzuki and I.\ Yotsuyanagi, Phys.Rev.D42 (1990)
4257, S.\ Hioki et.\,al. Phys.Lett.B272 (1991) 326, H.\ Shiba and T.\ Suzuki, Phys.Lett.B333 (1994) 461, J.\ D.\
Stack, S.D.\ Neiman and R.J.\  Wensley Phys.Rev.D50 (1994) 3399

\bibitem{zakharov} V.G.\ Bornyakov  et.\,al.;
 Phys.Lett.B537:291-296,2002 e-Print Archive:
hep-lat/0103032

\bibitem{zakharov1}  F.V.\ Gubarev and V.I.\ Zakharov
e-Print Archive: hep-lat/0204017

\bibitem{polikarpov} M.N.\ Chernodub et.\,al;
Nucl.Phys.Proc.Suppl.102:347-354,2001; T.\ Suzuki et.al; Nucl.Phys.Proc.Suppl.106:631-633,2002; e-Print Archive:
hep-lat/0110059

\bibitem{tomboulis1} T.G.\ Kovacs and E.T.\ Tomboulis, Phys.Rev.D65:074501,2002
e-Print Archive: hep-lat/0108017


\bibitem{gubarev} F.V.\ Gubarev; e-Print Archive:
hep-lat/0204018

\bibitem{deforcrand} P.\ de Forcrand and M.\ Pepe;
Nucl.Phys.B598:557-577,2001 e-Print Archive: hep-lat/0008016

\bibitem{vortexscaling} K.\ Langfeld, H.\ Reinhardt and O.\ Tennert;
Phys.Lett.B419:317-321,1998 e-Print Archive: hep-lat/9710068

\bibitem{laplacianvortex} K.\ Langfeld, H.\ Reinhardt and A.\ Schafke;
Phys.Lett.B504:338-344,2001 e-Print Archive: hep-lat/0101010


\bibitem{zakharov2} F.V.\ Gubarev and V.I.\ Zakharov; e-Print Archive: hep-lat/0211033







\end{thebibliography}
\end{document}